\documentstyle[12pt]{article}

\font\tenrm=cmr10

\newcommand{\ben}{\begin{equation}}
\newcommand{\een}{\end{equation}}
\newcommand{\bea}{\begin{eqnarray}}
\newcommand{\eea}{\end{eqnarray}}
\newcommand{\nn}{\nonumber \\ }
\newcommand{\bdm}{\begin{displaymath}}
\newcommand{\edm}{\end{displaymath}}

\newcommand{\hf}{\frac{1}{2}}
\newcommand{\eq}{\begin{equation}}
\newcommand{\en}{\end{equation}}
\newcommand{\eqn}{\begin{eqnarray}}
\newcommand{\enn}{\end{eqnarray}}
\newcommand{\CR}{\nonumber \\}
\newcommand{\refeq}[1]{Eq.(\ref{#1})}
\newcommand{\refta}[1]{(\ref{#1})}
\newcommand{\twt}{\frac{2}{3}}
\newcommand{\hl}{\\ \hline}
\newcommand{\ad}{{\rm ad}}
\newcommand{\BC}{{\bf C}}

\newcommand{\BG}{{\bf g}}
\newcommand{\Bn}{{\bf n}}
\newcommand{\Bh}{{\bf h}}
\newcommand{\cM}{{\cal M}}
\newcommand{\cF}{{\cal F}}
\newcommand{\cN}{{\cal N}}
\newcommand{\EXP}{{\rm exp}}
\newcommand{\E}{{\rm e}}
\newcommand{\I}{{\rm i}}
\newcommand{\half}{{1\over2}}
\newcommand{\HBG}{\hat{{\bf g}}}
\newcommand{\pa}{\partial}
\newcommand{\rank}{{\rm rank}}

\newcommand{\AT}[2]{{\scriptstyle #1}\atop{\scriptstyle #2}}
\newcommand{\A}{\alpha}
\newcommand{\B}{\beta}
\newcommand{\D}{\delta}
\newcommand{\DE}{\Delta}
\newcommand{\G}{\gamma}

\newcommand{\vep}{\varepsilon}

\newcommand{\T}{\theta}

\newcommand{\LM}{\Lambda}
\newcommand{\lm}{\lambda}

\newcommand{\vp}{\varphi}

\newcommand{\CMP}{{\it Commun. Math. Phys.}}

\newcommand{\IJMP}{{\it Int. J. Mod. Phys.}}
\newcommand{\NPB}[1]{{\it Nucl. Phys.} {\bf B#1}}
\newcommand{\PLB}[1]{{\it Phys. Lett.} {\bf B#1}}
\newcommand{\MPLA}[1]{{\it Mod. Phys. Lett.} {\bf A#1}}
\newcommand{\IJMPA}[1]{{\it Int. J. Mod. Phys.} {\bf A#1}}

\newcommand{\TMP}[1]{{\it Theor. Math. Phys.} {\bf #1}}

\newcommand{\AP}[1]{{\it Ann. Phys.} {\bf #1}}
\newcommand{\LNiM}[1]{{\it Lect. Notes in Math. vol.} {\bf #1}}

\newcommand{\JMP}[1]{{\it J. Math. Phys.} {\bf #1}}
\newcommand{\PTPS}[1]{{\it Prog. Theor. Phys. Suppl.} {\bf #1}}

\newcommand{\ASPM}[1]{{\it Adv. Stud. Pure. Math.} {\bf #1}}
\newcommand{\AM}[1]{{\it Adv. Math.} {\bf #1}}

\def\bA{\bar{\alpha}}
\def\bLM{\bar{\Lambda}}
\def\blm{\bar{\lambda}}

\newcommand{\dz}{{d z\over 2\pi\I}}
\def\ord#1#2{{#2\over (z-w)^{#1}}}
\def\ordo#1{{#1\over z-w}}

\def\sdim{{\rm sdim}}
\def\WG{\widehat{G}}

\def\Tb{\tilde{b}}
\def\Tc{\tilde{c}}
\def\TT{\tilde{T}}
\def\TG{\tilde{G}}

\begin{document}
\renewenvironment{thebibliography}[1]
  { \begin{list}{\arabic{enumi}.}
    {\usecounter{enumi} \setlength{\parsep}{0pt}
     \setlength{\itemsep}{3pt} \settowidth{\labelwidth}{#1.}
     \sloppy
    }}{\end{list}}

\parindent=1.5pc
\begin{flushright}
NBI-HE-92-81\\
October 1992
\end{flushright}
\vglue 0.6cm
\begin{center}{{\bf EXTENDED SUPERCONFORMAL ALGEBRAS\\
                \vglue 3pt
               FROM CLASSICAL AND QUANTUM \\
               \vglue 3pt
               HAMILTONIAN REDUCTION
\footnote{ To appear in the proceedings of the International
Workshop on ``String Theory, Quantum Gravity and the Unification of the
Fundamental Interactions", Rome, September 21-26, 1992; talk presented by
J.L. Petersen}}\\
\vglue 1.0cm
{KATSUSHI ITO, JENS OLE MADSEN}\\[.2cm]
{and}\\[.2cm]
{JENS LYNG PETERSEN}\\[.2cm]
{\it Niels Bohr Institute, University of Copenhagen,}\\
{\it Blegdamsvej 17, DK 2100 Copenhagen \O , Denmark}
\vglue 0.8cm
{ABSTRACT}}
\end{center}
\vglue 0.3cm
{\rightskip=3pc
 \leftskip=3pc
 \tenrm\baselineskip=12pt
 \noindent
The classification of extended superconformal algebras of the
Knizhnik-Bershadsky type with $W$-algebra like composite operators
occurring in the commutation relations, but with generators only of
conformal dimension $1,\frac{3}{2}$ and 2, has recently been
reconsidered by various authors from various points of view. We argue
that a particularly natural classification seems to arise on the basis
of hamiltonian reduction of affine Lie superalgebras with even
subalgebras $G\oplus sl(2)$. Based on generic formulas for the Poisson
bracket structure of the classical Gel'fand-Dickey algebra, we
introduce similar generic formulations for all these algebras at the
quantum level. Similarly, we rewrite the free field (Feigin-Fuchs)
representations of all these algebras by using the BRST formalism and
the free field realization of the affine Lie superalgebra. Again we
emphasize the unifying aspects partly based on previously known results,
partly on our own work when everything is formulated in the language
of reduced affine Lie superalgebras. We also discuss the screening
operators of these algebras. Although completely explicit results for
all of those are known only in a  limited number of cases,
enough may be said about the general structure, to in fact provide what
appears to be the complete pattern
of singular vectors in the free field realisation.
\vglue 0.8cm}

{\bf\noindent 1. Introduction}
\vglue 0.2cm
The role of extended superconformal algebras in formulations of
such related subjects as string theory, topological field theory and
conformal
field theory has been remarkable. This has been true both as far as the
world sheet supersymmetry and the target space geometry is concerned.
It seems not unlikely that new and unforeseen applications will turn up.
Thus it appears to be of general interest to have these extended
superconformal algebras reasonably classified.

One must distinguish several different ways in which the supersymmetries
have been implemented. The first way studied considered only what may be
termed extended superconformal {\em Lie} (or perhaps,
super-Lie) algebras: those for which commutators (anticommutators) of
generators are themselves members of the algebra, i.e., the expression for
the commutators involve only a {\em linear} combination of the
generators themselves. This class of algebras appears to have been
adequately classified for some time. For these algebras it seems
reasonable to restrict the number, $N$ of supersymmetries to be less
than 4.
{}For example, the linear $o(N)$ supersymmetric extension of the Virasoro
algebra, which was introduced by Ademollo et al.\cite{Ade}, was
realized as the superconformal transformation on extended superspace
with internal $o(N)$ symmetry. However, this formulation leads to
negative  conformal weight generators and the
absence of a central extension unless $N\leq 4$ \cite{Ade}, \cite{Sch}.

In a slightly different approach one regards linear $N=2,4$ extended
superconformal algebras as a symmetry of supersymmetric non-linear
sigma models on group manifolds or on  coset spaces such as hermitian
symmetric spaces \cite{KaSu} or Wolf spaces \cite{sigma}.

Knizhnik and Bershadsky \cite{KnBe}, however,  have introduced the
{\em non-linear} extension of superconformal algebras with
$u(N)$ and $so(N)$ affine symmetries. The precise form of the algebra
follows from closure and associativity
of the operator product expansions (the OPE method).
This non-linearity means that these extended superconformal algebras
are closely related to Zamolodchikov's $W$-algebras \cite{Zamo}.
$W$-algebras are known to play a very useful role in the classification
of chiral algebras in conformal field theories, so this non-linear
extension deserves further study. The representation theory has been
studied in the case of  $so(N)$-extended algebras \cite{Sc} and in
special cases such as the so-called doubly extended $N=4$ algebra
\cite{sigma,ivanov} for which a fair amount by now is known\cite{dob}.

{}For actual applications in conformal field theory or string theory,
it must be expected that the extended superconformal algebra will be
a sub-algebra of an even larger, perhaps $W$-extended (or super
$W$-extended) chiral  algebra. However, it seems of some interest
to first address the classification problem for the extended
superconformal algebras themselves, especially since that problem
appears so much simpler. The classification presented here has
independently been found by Bowcock and by Fradkin and
Linetsky\cite{BoFrLi,FrLi}.  Our emphasis here is on the unifying
aspect provided by  the idea of hamiltonian reduction of Lie
superalgebras.

As is by now well known, there is a beautiful way of obtaining
$W$-algebras from the hamiltonian reduction
\cite{DrSo,Be,BeOo,BaFeFoRaWi} of suitable affine Lie algebras by
considering the constraints on the phase space of currents using a
gauge fixing procedure, (see also the recent review by Bouwknegt and
Schoutens \cite{BoSc}). Recently various types of further extensions
of the $W$-algebras \cite{Po,Bers,FeOrRuTsWi,BaTjDr,Rom}
including their supersymmetric generalization \cite{Kh,BeOo2,superW} have
been considered, also by using the hamiltonian reduction technique.

There is a simple rule for finding at least the conformal dimension of the
$W$-generators in any one particular case. In fact the labelling of the
$W$-algebra as proposed by the review of Bouwknegt and
Schoutens\cite{BoSc}  makes use of that directly. The
reduction is characterized by a particular $sl(2)$ subalgebra of the
original Lie algebra. The generators of the original Lie algebra
occur as spin multiplets of that $sl(2)$ subalgebra with spins
$\{s_i\}$, say. Then the conformal dimensions of the generators of
the ensuing $W$-algebra are simply the set of numbers, $\{s_i+1\}$,
the constant 1, being the conformal dimension of the WZNW currents of
the unconstrained affine lie-algebra.

In our case we want to consider the set of extended superconformal
algebras, which may be characterized as follows: (i)
There is an energy-momentum tensor of conformal dimension 2, (ii) there
is a number of supercurrents, $G_{\G}$, primary conformal fields of
dimension $\frac{3}{2}$ and primary affine fields of some affine algebra
$\hat{G}$, transforming according to some completely reducible
representation of $G$, and where the label $\G$, is a weight label of
that representation, or indeed, as we shall see, an odd root of the
Lie superalgebra; (iii) there are the generators of the affine Lie
algebra $\hat{G}$, with conformal dimension 1; (iv) all other generators
are composites of the above.

It follows that within the framework of hamiltonian reduction of Lie
superalgebras which we are going to adopt, we need consider situations
where a certain $sl(2)$ subalgebra is embedded so that the generators
of the subalgebra, $G$, have spin 0. That immediately implies that we
are looking for situations where there is a subalgebra of the form
$G\oplus sl(2)$. The generators of the $sl(2)$ itself have spin 1 and
indeed give rise as usual to the energy momentum tensor after reduction.
We see that we are looking for situations where the remaining
generators have spin $\hf$.

It is easy from the known theory of Lie superalgebras \cite{Kac}
to look up the ones having even subalgebras of the form $G\oplus sl(2)$.
The result is given in table \refta{ta:li}. Previously we \cite{ItMaPe}
have considered these in some detail after two of us considered the
classical reduced algebras \cite{ItMa}.
\begin{table}
\caption{Lie Superalgebras with an even subalgebra $A_{1}$}
\label{ta:li}
\begin{center}
\begin{tabular}{|lllll|}                         \hline\hline
$\BG$ & $\BG$ &    $\BG_{0}$      & $G$ & $G$ (alternatively)    \\ \hline
$A(n|1)$ & $sl(n+1,2)$
& $A_{n}\oplus A_{1}\oplus u(1)$  & $A_{n}\oplus u(1)$ &
$sl(n+1)\oplus u(1)$ \\
$B(n|1)$ & $osp(2n+1|2)$  & $B_{n}\oplus A_{1}$  & $B_{n}$ & $so(2n+1)$\\
$D(n|1)$ & $osp(2n|2)$ & $D_{n}\oplus A_{1}$  & $D_{n}$ & $so(2n)$\\
$D(2|n)$ & $osp(4|2n)$
& $A_{1}\oplus A_{1}\oplus C_{n}$ & $A_{1}\oplus C_{n}$ &
$sp(2n)\oplus sl(2)$\\
$D(2|1;\A)$ & & $A_{1}\oplus A_{1}\oplus A_{1}$
& $A_{1}\oplus A_{1}$ & $sl(2)\oplus sl(2)$ \\
$F(4)$  && $B_{3}\oplus A_{1}$  & $B_{3}$ & $so(7)$ (spin$(7)$)\\
$G(3)$    &   & $G_{2}\oplus A_{1}$ & $G_{2}$ &$G_2$ \\ \hline
$B(1|n)$  & $osp(3|2n)$ &
$A_{1}\oplus C_{n}$  & $C_{n}$ & $sp(2n)$\\ \hline\hline
\end{tabular}
\end{center}
\end{table}
The well known superconformal algebras occur as special cases. Thus $N=1$
corresponds to $osp(1|2)$ and $N=2$ to $osp(2|2)$ as is very well known.
{}For $N=4$, there are several cases with $D(2|1)$ and $A(1|1)$
forming extremes between which $D(2|1;\A)$ interpolates.
It is amusing and perhaps significant to notice that this
classification corresponds to that of the reduced holonomy
groups of non-symmetric Riemannian manifolds \cite{Ber}.
In nearly all of the cases in table \refta{ta:li} the odd generators
do indeed carry spin $\hf$ of the $sl(2)$ subalgebra. There is only
one exception: in the case of $B(1|n)$ the spin is 1, so that the
supercurrents acquire conformal dimension 2 after reduction. Hence
this kind of algebra falls outside the scope of what we consider here,
but may deserve separate investigation. In the appendix we provide
details on the root systems of these algebras.

We expect that these extended superconformal algebras might correspond to
non-linear sigma models with rich geometrical structures such as
quaternionic or octonionic structures.
In the similar classification based mostly on the OPE method\cite{BoFrLi}
the role of an underlying Lie superalgebra was also hinted at in various ways.
Our treatment presented here, however, provides a complete account also of
the free field realizations.

To avoid confusion, is should be emphasized that the affine Lie superalgebras
we are going to consider differ from the supersymmetrization of WZNW models
\cite{DiKnPeRo} giving rise to super affine Lie algebras. However, it may be
of interest to consider the two ideas put together, as has
recently been done\cite{FrRaSo}. The results have a slightly different
perspective and seems of particular value when further $W$ type extensions
need be considered. Thus one considers the embedding, not of $sl(2)$, but
of the ``supersymmetric extension", $osp(1|2)$. But that does not seem
necessary for our purpose of understanding the extended superalgebras
per se, without further extensions.

The purpose of the present paper is to present the hamiltonian
reduction in the classical and quantum cases and to provide a unifying
account of all these extended superconformal algebras. Similarly we
study the quantum algebras further using free field representations,
emphasizing again the unifying aspects offered by the formalism, so that
generic formulas may be written down. For practical calculations the
free field representations are
useful for the computation of correlation functions once the pattern of
singular vectors have been understood \cite{FeFu}. To that end we analyze
the structure of screening operators.

This paper is organized as follows:
In sect. 2, we first review some basic properties of affine Lie
superalgebras, and then we discuss the classical hamiltonian reduction
for an affine Lie superalgebra $\HBG$ associated with a Lie
superalgebra $\BG$ which has the even subalgebra $G\oplus sl(2)$, and we
derive the classical $G$ extended superconformal Gel'fand-Dickey algebra.
In sect. 3 we use the BRST gauge fixing procedure and the
Wakimoto realization \cite{FeFr} of the affine Lie superalgebra to derive
the free field realization of the extended superconformal algebra at the
level of BRST charges and energy momentum tensors. In sect 4, we explain
how the classical results for the algebras generalize in all cases
and we complete the free field construction with generic formulas pertaining
to all cases. In sect. 5, we investigate the structure of screening
operators and the null field structure of degenerate representations of
the $G$ extended superconformal algebras.
\vglue 0.6cm

{\bf\noindent 2. The Classical Hamiltonian Reduction}
\vglue 0.2cm
{\it\noindent 2.1 Lie superalgebras}
\vglue 0.1cm
We start with explaining some definitions of basic classical Lie
superalgebras \cite{Kac} and their affine extensions.
Let $\BG=\BG_{0}\oplus\BG_{1}$ be a rank $n$ basic classical Lie
superalgebra with an even subalgebra $\BG_{0}$ and an odd subspace
$\BG_{1}$. $\DE=\DE^{0}\cup\DE^{1}$ is the set of roots of $\BG$, where
$\DE^{0}$ ($\DE^{1}$) is the set of even (odd) roots. Denote  the set of
positive even (odd) roots by $\DE^{0}_{+}$ ($\DE^{1}_{+}$).
The superalgebra $\BG$ has a canonical basis
$\{ E_{\A}, e_{\G}, h^{i} \}$ ($\A\in\DE^{0}$, $\G\in\DE^{1}$,
$i=1, \ldots ,n$), which satisfies (anti-)commutation relations
\eqn
\mbox{[} E_{\A}, E_{\B} \mbox{]}&=&
\left\{ \begin{array}{ll}
N_{\A,\B} E_{\A+\B}, &\quad \mbox{ for $\A$, $\B$, $\A+\B\in \DE^{0}$},\\
\displaystyle{{2\A\cdot h \over \A^{2}}}, &\quad \mbox{ for $\A+\B=0$, }
\end{array}\right.  \CR
\left\{ e_{\G}, e_{\G'} \right\}&=& N_{\G,\G'} E_{\G+\G'}, \quad
\mbox{ for $\G, \G'\in \DE^{1}$, $\G+\G'\in \DE^{0}$},  \CR
\left\{ e_{\G}, e_{-\G} \right\}&=& \G\cdot h ,
\quad \mbox{for $\G\in\DE^{1}_{+}$}, \CR
\mbox{[} e_{\G}, E_{\A} \mbox{]}&=& N_{\G,\A} e_{\G+\A}, \quad
\mbox{ for $\A\in \DE^{0}$, $\G$,$\G+\A\in \DE^{1}$},  \CR
\mbox{[} h^{i}, E_{\A} \mbox{]}&=&\A^{i}E_{\A}, \quad
\mbox{[} h^{i}, e_{\G} \mbox{]}=\G^{i}e_{\G}.
\enn
The even subalgebra $\BG_{0}$ is generated by $\{ E_{\A}, h^{i} \}$.
The odd subspace $\BG_{1}$ is spanned by $\{e_{\G}\}$.
$\BG_{0}$ acts on $\BG_{1}$ as a faithful, completely reducible
representation. The Killing form $( \ ,\ )$ on $\BG$ is defined by
\eq
(E_{\A}, E_{\B})={2\over \A^{2}}\D_{\A+\B,0}, \quad
(e_{\G}, e_{-\G'})=-(e_{-\G}, e_{\G'})=\D_{\G,\G'},  \quad
(h^{i},h^{j})=\D_{i j},
\en
for $\A,\B\in\DE^{0}$, $\G,\G' \in \DE^{1}_{+}$, $i,j=1,\ldots, n$.

An affine Lie superalgebra $\HBG$ at level $k$ is the
(untwisted) central extension of $\BG$ and consists of the elements
of the form $(X(z), x_{0})$, where $X(z)$ is a $\BG$-valued
Laurent polynomial of $z\in \BC$ and $x_{0}$ is a number
\cite{reimansemenov}. The commutation relation for two elements
$(X(z), x_{0})$ and $(Y(z), y_{0})$ is given by
\eq
\mbox{[} \ (X(z), x_{0}), \ (Y(z), y_{0}) \ \mbox{]}
=(\ \mbox{[}\ X(z), \ Y(z) \ \mbox{]},\ k \oint \dz (X(z), \pa Y(z))\ ).
\label{eq:com}
\en
The dual space $\HBG^{*}$ of $\HBG$ is generated by the current
$(J(z), a_{0})$. Using the Killing form $( \ , \ )$ on $\BG$, we may
identify $\HBG^{*}$ with $\HBG$. The inner product
$\langle\ , \ \rangle$ of $(J(z), a_{0}) \in\HBG^{*}$
and $(X(z), x_{0}) \in\HBG$ is given by
\eq
\langle \ (J , a_{0}), \ (X, x_{0}) \ \rangle
=\oint\dz (\ J(z),\  X(z)\ ) +a_{0}x_{0}.
\en
One defines the coadjoint action $\ad^{*}$ of $\HBG$
on $\HBG^{*}$ by
\eq
\langle\ \ad^{*}(X, x_{0})\ (J, a_{0}),\ (Y, y_{0})\ \rangle
=-\langle\ (J, a_{0}),\
\mbox{[}\ (X, x_{0}),\ (Y, y_{0})\ \mbox{]}\ \rangle.
\en
Using \refeq{eq:com}, one gets
\eq
\ad^{*}(X, x_{0})(J(z), a_{0})
=(\ \mbox{[}\ X(z), J(z)\ \mbox{]}+k a_{0}\pa X(z),\ 0\ ).
\en
Namely, the coadjoint action simply takes the form of an infinitesimal gauge
transformation of the current $J(z)$.
Denote this gauge transformation with the gauge parameter $\LM(z)$
as $\D_{\LM}$:
\eq
\D_{\LM} J(z)=\mbox{[} \ \LM(z),\  J(z)\ \mbox{]} +k a_{0} \pa \LM (z),
\label{eq:gauge1}
\en
In the following we take the number $a_{0}$ to be 1.
In terms of the canonical basis
\eq
J(z)=\sum_{\A\in\DE^{0}}{\A^{2}\over 2}J_{\A}(z)E_{\A}
     +\sum_{\G\in\DE^{1}}j_{\G}(z)e_{\G}+\sum_{i=1}^{n}H^{i}(z)h^{i}.
\label{eq:current}
\en
and
\eq
\LM(z)=\sum_{\A\in\DE^{0}}\vep_{\A}(z)E_{\A}
     +\sum_{\G\in\DE^{1}}\xi_{\G}(z)e_{\G}
     +\sum_{i=1}^{n}\vep^{i}(z)h^{i},
\en
we can express the gauge transformation \refeq{eq:gauge1} in terms
of components:
\eqn
\D_{\LM} J_{\A}\!\!\!\! &=& \!\!\!\! \sum_{\B, \A-\B\in \DE^{0}}
             {(\A-\B)^{2}\over \A^{2}}N_{\B,\A-\B} \vep_{\B}J_{\A-\B}
            -{2 (\A\cdot H) \over \A^{2}} \vep_{\A}
            +{2k\over \A^{2}}\pa\vep_{\A}
            +(\A\cdot\vep) J_{\A} \CR
         & & +\sum_{\G,\A-\G\in \DE^{1}}
             {2\over \A^{2}}N_{\G,\A-\G} \xi_{\G}j_{\A-\G}, \CR
\D_{\LM} j_{\G} \!\!\!\! &=& \!\!\!\!
             \sum_{\AT{\A\in\DE^{0}}{\G-\A\in\DE^{1}}}
             \!\!\!\! N_{\A,\G-\A}\vep_{\A}j_{\G-\A}
            +(\G\cdot\vep)j_{\G}
             +\!\!\!\! \sum_{\AT{\A\in\DE^{0}}{\G-\A\in\DE^{1}}}
       \!\!\!\! {\A^{2}\over 2}N_{\G-\A,\A}\xi_{\G-\A}J_{\A}
             -\xi_{\G} \G\cdot H +k\pa\xi_{\G}, \CR
\D_{\LM} H^{i} \!\!\!\! &=& \!\!\!\!
             \sum_{\A\in\DE^{0}}\A^{i}\vep_{\A}J_{-\A}
             +\sum_{\G\in\DE^{1}_{+}}
              \G^{i}(\xi_{\G}j_{-\G}+\xi_{-\G}j_{\G})
             +k\pa\vep^{i}.
\label{eq:gauge2}
\enn
Writing the gauge transformations $\D_{\LM}$ as
\eqn
\D_{\LM}&=&\oint\dz (\LM(z), J(z)) \CR
        &=&\oint\dz \sum_{\A\in\DE^{0}}\vep_{\A}J_{-\A}
           +\sum_{\G\in\DE^{1}_{+}}(j_{\G}\xi_{-\G}+\xi_{\G}j_{-\G})
           +\sum_{i=1}^{n}\vep^{i}H^{i},
\enn
one can introduce a canonical Poisson structure on the dual space
$\HBG^{*}$.
This Poisson structure is conveniently summarized in the form of ``the
operator product expansions": Poisson brackets between modes are obtained
from these exactly as commutators from OPE's in conformal field theory.
The result is:
\eqn
J_{\A}(z)J_{\B}(w)&=&\left\{
\begin{array}{ll}
\displaystyle{\ordo{N_{\A,\B}J_{\A+\B}(w)}+\cdots} ,& \quad
\mbox{ for $\A$, $\B$, $\A+\B\in \DE^{0}$}, \\
 \displaystyle{\ord{2}{{2k\over \A^{2}}}
 +\ordo{ {2\A\cdot H(w)\over \A^{2}} }+\cdots}, &\quad
\mbox{ for $\A+\B=0$,}
\end{array}\right. \CR
j_{\pm\G}(z)j_{\pm\G'}(w)&=&
\displaystyle{+\ordo{N_{\mp\G,\mp\G'}J_{\pm(\G+\G')}(w)}+\cdots} , \quad
\mbox{ for $\G,\G'\in\DE^{1}_+$, $\G+\G'\in\DE^{0}$}, \CR
j_{\pm\G}(z)j_{\mp\G'}(w)&=&
\displaystyle{-\ordo{N_{\mp\G,\pm\G'}J_{\pm(\G-\G')}(w)}+\cdots} , \quad
\mbox{ for $\G,\G'\in\DE^{1}_+$, $\G-\G'\in\DE^{0}$}, \CR
j_{\G}(z)j_{-\G}(w)&=&-\ord{2}{k}-\ordo{\G\cdot H(w)}+\cdots,
\quad \mbox{ for $\G\in\DE^{1}_{+}$,}\CR
J_{\A}(z)j_{\G}(w)&=&{N_{-\A,\G +\A}j_{\G+\A}(w)\over z-w}+\cdots ,\quad
\mbox{ for $\G , \G+\A\in \DE^{1}$, $\A\in \DE^{0}$, },  \CR
H^{i}(z)J_{\A}(w)&=&{\A^{i}J_{\A}(w)\over z-w}+\cdots , \quad
H^{i}(z)j_{\G}(w)={\G^{i}j_{\G}(w)\over z-w}+\cdots. \CR
H^{i}(z)H^{j}(w)&=&{k\D^{i j}\over (z-w)^{2}}+\cdots .
\enn
Here we have used some identities for the structure constants coming
from the Jacobi identities:
\eqn
{2N_{\A,\B}\over (\A+\B)^{2}}&=&{2 N_{\B,-\A-\B}\over \A^{2}}
                               ={2 N_{-\A-\B,\A}\over \B^{2}},
\quad \mbox{for $\A,\B \in\DE^{0}$, }\CR
{2N_{\G,\G'}\over (\G+\G')^{2}}&=&
N_{\G',-\G-\G'}=-N_{-\G-\G',\G}, \quad\mbox{for $\G,\G'\in\DE^{1}_{+}$ }\CR
{2N_{\G,-\G'}\over (\G-\G')^{2}}&=&
N_{-\G',-\G+\G'}=N_{-\G+\G',\G}, \quad\mbox{for $\G,\G'\in\DE^{1}_{+}$ }\CR
{2N_{-\G,-\G'}\over (\G+\G')^{2}}&=&
-N_{-\G',\G+\G'}=N_{\G+\G',-\G}, \quad\mbox{for $\G,\G'\in\DE^{1}_{+}$ }.
\label{eq:str}
\enn

In the following, as argued in the introduction,
we will study the class of Lie superalgebras,
with even algebras of the form $G\oplus A_{1}$.
Using Kac's notation, these algebras are classified as follows
$A(n|1)\ (n\geq 1)$, $A(1,0)=C(2)$, $B(n|1)\ (n\geq 0)$,
$B(1|n)\ (n\geq 1)$, $D(n|1)\ (n\geq 2)$, $D(2|n)\ (n\geq 1)$,
$D(2|1;\A)$, $F(4)$ and $G(3)$ (see table \ref{ta:li}).
The embedding of $A_{1}$ in $\BG$ carried by ${\bf g_1}$
has spin $\half$, except for $B(1|n)$.
In the case of $B(1|n)$ the embedding has spin 1.
In the spin $\half$ embedding case, which is the only case we will be
studying in this paper for the reasons given in the introduction,
the odd subspace $\BG_{1}$ belongs to the spin $\half$ representation
with respect to the even subalgebra $A_{1}$ and therefore
splits into two parts:
\eq
\BG_{1}=(\BG_{1})_{+\half}\oplus(\BG_{1})_{-\half}.
\label{eq:dec}
\en
Likewise, the odd root space $\DE^{1}$
may be divided into two parts
\eq
\DE^{1}=\DE^{1}_{+\half}\cup\DE^{1}_{-\half},
\en
where $(\BG_{1})_{\pm\half}$ is spanned by the generators whose roots
belong to $\DE^{1}_{\pm \half}$.
More explicitly, the sets $\DE^{1}_{\pm \half}$ consist of the roots
$\G\in \DE^{1}$ satisfying
\eq
{\G \cdot \theta \over \theta^{2}}=\pm\half,
\label{eq:odd1}
\en
where $\theta$ is the simple root of $A_{1}$.
By choosing appropriate simple roots for $\BG$, we can take the root
space $\DE^{1}_{+\half}$  as the space of odd positive roots;
\eq
\DE^{1}_{+}=\DE^{1}_{+\half},
\en
and then
\eq
\DE^{1}_{+\half}=\DE^{1}_{-\half}+\theta,
\label{eq:odd2}
\en
holds.
We note that each odd space $(\BG_{1})_{\pm\half}$ belongs to a
fundamental representation of $G$ of dimension $|\DE^{1}_{+}|$, this
being the number of positive odd roots. These will be related to
the supercurrents as discussed below.
The corresponding  representation of $G$ (possibly of the form
$G^{(1)}\oplus G^{(2)}$)has
weights simply equal to
$$\G_{\perp}(G^{(i)})\equiv P_{G^{(i)}}(\G)$$
namely the (positive in this case)
odd root projected onto the root space of $G^{(i)}$. If we denote the
representation matrices in the canonical basis by
$$t^{\A}_{\G ,\G '},t^i_{\G ,\G '}$$
then we have (using the Jacobi identities, see also below)
for $\G\in\DE^1_+$
\bea
t^{\A}_{\G ,\G '}&=&-N_{-\A ,-\G }\D_{\G+\A ,\G'}\nn
t^i_{\G,\G'}&=&(\G_{\perp})^i\D_{\G,\G'}\nn
tr(t^it^j)&=&c_F\D_{ij}\nn
tr(t^{\A}t^{\B})&=&c_F\frac{2}{\A^2}\D_{\A+\B,0}\nn
c_F&=&\sum_{\G\in\DE^1_+}\G_{\perp}^2/r(G)
\eea
where
$$r(G)=\mbox{rank}(G).$$
Table (\ref{ta:liprop}) gives several parameters relevant for the different
Lie super algebras we are considering.
\begin{table}
\caption{Properties of Lie superalgebras giving rise to ESA's}
\label{ta:liprop}
\begin{center}
\begin{tabular}{|lllllrr|}                         \hline\hline
$\BG$        & \sdim &rank &$h^{\vee}$  & $|\DE^1_+|$      & $c_F^{(1)}$
&$c_F^{(2)}$     \\ \hline
$A(n|1)$     & $n^{2}-2n$ ($-2$ for $n=1$) & $n+2$ &$n-1$
& $2(n+1)$  & $2$& $\frac{(n-1)^2}{2(n+1)}$ \\
$B(n|1)$     & $(n-1)(2n-1)$ &$n+1$ &$3-2n$
& $2n+1$  & $-2$& \\
$D(n|1)$     & $(n-1)(2n-3)$ &$n+1$ &$4-2n$
& $2n$  & $-2$& \\
$D(2|n)$     & $(n-2)(2n-3)$ & $n+2$& $2n-2$
& $4n$ & $4$ & $-2n$ \\
$D(2|1;\A)$  & $1$ &$3$ &$0$  & $4$
& $-2\G$ &$-2(1-\G)$ \\
$F(4)$       & $8$ &$4$ &$-3$  & $8$  & $-2$& \\
$G(3)$       & $3$ &$3$ &$2$  & $7$ & $2$ & \\ \hline\hline
\end{tabular}
\end{center}
\end{table}
One may show that
$$t^{\A}_{\G,\G'}=-t^{\A}_{\T-\G',\T-\G}.$$
The Casimir operator in that representation takes the form
\ben
C_{\G,\G'}=(\sum_{\A\in\Delta^0(G)}\frac{\A^2}{2}t^{\A}t^{-\A}+
\sum_{i=1}^{r(G)} t^it^i)_{\G,\G'}
\een
Then it is convenient to introduce the currents corresponding to $\WG$
in the following basis
\ben
J_{\G,\G'}(z)\equiv\sum_{\A\in\Delta^0(G)}\frac{\A^2}{2}t^{\A}_{\G,\G'}
J_{-\A}(z)+\sum_{i=1}^{r(G)}t^i_{\G,\G'}H_i(z)
\label{eq:JGGp}
\een
Notice that for given $\G,\G'$, $J_{\G,\G'}$ contains at most a single term,
which may be written in several ways, for example as:
\ben
J_{\G,\G'}(z)=\left\{
\begin{array}{ll}
N_{\G',-\G}J_{\G -\G'}(z)&\mbox{for \ \ }\G -\G'\in\DE^0(G)\\
&\\
\G\cdot H(z)&\mbox{for \ \ }\G=\G'\\
&\\
0&\mbox{otherwise}
\end{array}\right.
\een

It follows that the root system of the even subalgebra of
$G$ is expressed in the form of $\G-\G'$, where $\G$ and $\G'$ are
some positive odd roots.
This fact turns out to be essential for the study of
the structure of general $G$
extended superconformal algebras.
\vglue 0.2cm
{\it\noindent 2.2 Classical hamiltonian reduction}
\vglue 0.1cm
We now consider the hamiltonian reduction of the phase space of the
currents $\HBG^{*}$ associated with any one of the set of
Lie superalgebras $\BG$
with an even subalgebra $G\oplus A_{1}$.
We want to impose the constraint on the even subalgebra
$A_{1}$ while keeping the affine $G$ algebra symmetry.
We start from  the phase space $\cF$ of the currents with the constraint
\eq
J_{-\theta}(z)=1 .
\label{eq:cons1}
\en
Denote by $\cN$ the gauge group which preserves the constraint
\refeq{eq:cons1}.
Putting $-\theta$ for $\A$ in the first formula of eqs. \refeq{eq:gauge2},
we get
\eq
\D_{\LM} J_{-\theta}=
            {2 (\theta\cdot H) \over \theta^{2}} \vep_{-\theta}
            +{2k\over \theta^{2}}\pa\vep_{-\theta}
            -(\theta\cdot\vep) J_{-\theta}
         +\sum_{\G \in \DE^{1}_{+}}
     {2\over \theta^{2}}N_{-\G,-\theta+\G} \xi_{-\G}j_{-\theta+\G}.
\label{eq:gau1}
\en
Here we use the relation \refeq{eq:odd2} and $N_{\A, -\theta-\A}=0$
for  $\A\in\DE(G)$.
Thus $\cN$ is characterized by the algebra conditions
\eq
\vep_{-\theta}=\theta\cdot\vep=0, \quad
\xi_{-\G}=0, \quad {\rm for } \ \G\in\DE^{1}_{+}.
\en
So the Lie algebra $\Bn$ of the gauge group $\cN$ is
equal to $\WG\oplus \Bn_{1}$, where $\WG$ is
the affine Lie algebra corresponding to the even subalgebra $G$ and
$\Bn_{1}$ is generated by the elements:
\eq
\LM(z)=\vep_{\theta}(z)E_{\theta}
     +\sum_{\G\in\DE^{1}_{+}}\xi_{\G}(z)e_{\G}.
\en
Let $\cN_{1}$ be the gauge group corresponding to $\Bn_{1}$.
The reduced phase space $\cM$ is defined as the quotient space
$\cM=\cF/\cN_{1}$.
By the standard gauge fixing procedure, one chooses one or the other gauge
slice as a representative of the reduced phase space.
In  the so-called \lq\lq Drinfeld-Sokolov (DS)"-gauge the generators of the
classical extended algebra are rather easily read off. It is defined by
\cite{DrSo}:
\eq
J_{\theta}(z)=T (z), \quad
\theta \cdot H(z)=0, \quad
j_{\G}(z)=G_{\G}(z), \quad
j_{-\G}(z)=0,
\label{eq:dsgauge}
\en
for $\G\in\DE^{1}_{+}$.
And the generic gauge transformation projected on the DS gauge slice
becomes the transformation corresponding to the extended superconformal
algebra
\cite{DrSo,Be,BeOo,BaFeFoRaWi,Po,Bers,FeOrRuTsWi,BaTjDr,Rom,Kh,BeOo2}
in our case.
Hence the final step is to consider the generic gauge transformation
preserving the gauge condition \refeq{eq:dsgauge}. In particular
\eqn
\D_{\LM}J_{-\T} &=& -(\T\cdot\vep)
      +{2k\over \T^{2}}\pa\vep_{-\T}, \CR
\D_{\LM}(\T\cdot H) &=&
      \T^{2}\vep_{\T}-\T^{2}\vep_{-\T} T
      +k\pa(\T\cdot\vep)
      +\sum_{\G\in\DE^{1}_{+}}\T\cdot\G\xi_{-\G}G_{\G}, \CR
\D_{\LM}j_{-\G}&=&
      N_{-\T,\T-\G}\vep_{-\T}G_{\T-\G}
      +\sum_{\AT{\G'\in\DE^{1}_{+}}{\G'-\G\in\DE(G)}}
        N_{\G,-\G'}\xi_{-\G'}J_{\G'-\G}\CR
      &+&N_{\T-\G,\G}\xi_{\T-\G}+\xi_{-\G}(\G\cdot H)+k\pa\xi_{-\G},
\enn
for $\G\in\DE^{1}_{+}$.
By solving the conditions which preserve the DS-gauge
$\D J_{-\T}=0$, $\D (\T\cdot H)=0$ and $\D j_{-\G}=0$,
we can express the parameters $\vep_{\theta}$, $\theta\cdot\vep$ and
$\xi_{\G}$ ($\G\in\DE^{1}_{+}$) in terms of the
other parameters, $\vep_{-\theta}$, $\xi_{-\G}$ ($\G\in\DE^{1}_{+}$),
$\vep_{\A}$ and $\A\cdot \vep$ ($\A \in \DE(G)$):
\eqn
\T\cdot\vep\!\!\!\!&=& \!\!\!\! {2k\over \T^{2}}\pa\vep_{-\T},
\quad \quad \quad
\vep_{\T}=\vep_{-\T} T
          -{2k^{2}\over (\T^{2})^{2}}\pa^{2}\vep_{-\T}
          -\half \sum_{\G\in\DE^{1}_{+}}\xi_{-\G}G_{\G},
\label{eq:gaugetr1} \\
\xi_{\T-\G}\!\!\!\!&=& \!\!\!\!
           {-1\over N_{\T-\G,\G}}\bigl\{
           N_{-\T,\T-\G}G_{\T-\G}\vep_{-\T}
           +\!\!\!\! \sum_{\AT{\G'\in\DE^{1}_{+}}{\G-\G'\in\DE(G)}}
            \!\!\!\!N_{\G,-\G'}\xi_{-\G'}J_{\G'-\G}
           +\xi_{-\G}(\G\cdot H)+k\pa\xi_{-\G} \bigr\}, \nonumber
\enn
In the DS gauge,
the gauge transformation of $J_{\theta}(z)=T(z), J_{\B}(z), \B\cdot H(z)$
and $G_{\G}(z)$ may then be worked out in terms of these.
We refer to our papers\cite{ItMaPe}
for the straightforward but slightly messy details.

We then convert this result into formal ``operator product expansions".
We denote the gauge transformation in the DS-gauge as
\eq
\D=\oint\dz (\LM(z)\ ,J_{DS}(z)),
\en
where
\eq
J_{DS}(z)= \frac{\T^2}{2}T(z)E_{\theta}
          +\sum_{\G\in \DE^{1}_{+}}G_{\G}(z)e_{\G}
          +\sum_{\A\in\DE(G)}\frac{\A^2}{2} J_{\A}(z) E_{\A}
          +\sum_{i=1}^{\rank (G)}H^{i}h^{i},
\en
This conveniently generates
the Poisson bracket structure on the reduced phase space.
Here we take $h^{i}$ ($i=1, \ldots, \rank (G)$) as the generators of the
Cartan subalgebra of $G$.
We also introduce the rescalings $\TT={\T^{2}\over 2k}T$ and
$\TG_{\G}=\sqrt{{N_{\T-\G,\G}\over k}}G_{\G}$ for
$\G\in\DE^{1}_{+}$.
We further use several identities for the structure constants, like the
Jacobi identities \refeq{eq:str}, as well as
\eq
{N_{\T-\G,\G'-\T}\over N_{\G,\T-\G}}=-{N_{\G',-\G}\over N_{\G',\T-\G'}},
\en
for $\G,\G'\in\DE^{1}_{+}$,
as may be checked using explicit matrix representations
of $\BG$ \cite{Cor,DeNi} in table \ref{ta:li}.
Similarly one may derive
\ben
N_{\G ,\T -\G}=-N_{\G -\T ,-\G}=\frac{\T ^2}{2}
\een
from the fact that $\BG =G\oplus A_1$.
Using all these properties one may achieve a major simplification of the
complicated, generic expression we have previously presented\cite{ItMaPe}.
To this end it is very useful to make use of the affine currents of
$G$ in the basis of \refeq{eq:JGGp}. The result finally is:
\eqn
\TT(z)\TT(w)&=&\ord{4}{{-6k\over \T^{2}}}
         +\ord{2}{2\TT(w)}
         +\ordo{\pa \TT(w)}+\cdots, \CR
\TT(z)\TG_{\G}(w)&=&\ord{2}{{3\over2}\TG_{\G}(w)}+\ordo{\pa\TG_{\G}(w)}
     -\frac{1}{k}\frac{\sum_{\G'\in\Delta^1_+}J_{\G,\G'}G_{\G'}(w)}{z-w}\CR
J_{\B}(z)\TG_{\G}(w)&=&\ordo{-N_{-\B,-\G}\TG_{\B+\G}(w)}+\cdots, \quad
H^{i}(z)\TG_{\G}(w)=\ordo{\G^{i}\TG_{\G}(w)}+\cdots, \CR
\TG_{\T-\G'}(z)\TG_{\G}(w)&=&\ord{3}{2 k\D_{\G',\G}}
-2\frac{J_{\G,\G'}(w)}{(z-w)^2}\CR
&+&\frac{\D_{\G ,\G'}\frac{\T^2}{2}\TT-\pa J_{\G,\G'}(w)
+\frac{1}{k}(J^2)_{\G,\G'}(w)}{z-w}
+\cdots,
\label{eq:clope}
\enn
Here we imply the notation:
\ben
(J^2)_{\G,\G'}\equiv\sum_{\G''\in\DE^1_+}J_{\G,\G''}J_{\G'',\G}
\een
Notice that this expression is nonvanishing only if either $\G-\G'\in\DE(G)$,
$\G =\G'$ or $\G+\G'=\T$.
At this point,
the supercurrents $\TG_{\G}$ are not primary fields with respect
to $\TT(z)$.
However, if we define the total energy-momentum tensor
$T_{ESA}$ by adding the Sugawara form $T_{Sugawara}$ of the
$\WG$ affine Lie algebra: $T_{ESA}=\TT+T_{Sugawara}$ where
\bea
T_{Sugawara}&=&{1\over 2k}\left\{
\sum_{\A\in\DE(G)} {\A^{2}\over 2}J_{\A}J_{-\A}
     +\sum_{i=1}^{\rank (G)}H^{i}H^{i} \right\}\nn
&=&\frac{1}{2kc_F}tr(J^2),
\eea
we may check that the supercurrents have conformal weight 3/2 with
respect to $T_{ESA}$. The result is the classical or Gel'fand-Dickey
extended superconformal algebra. In sect. 4 we shall give the explicit form
generalized to the arbitrary quantum case.
The classical value of the central charge $c_{ESA}$ is
$-12k/\T^{2}$. Similarly, the classical central extension of the surviving
affine Lie algebra $\WG$ (or $\WG=\oplus_i\WG^{(1=i)}$
is given by $K^{(i)}_{cl}=\frac{2k}{\A_{Li}^2}$,where $\A_L$ denotes the
``long root" so that this definition is the one pertaining to a normalization
where the longest root has length squared equal to 2.
The relative minus sign between
these expressions is significant. In fact the metrics in the root space of the
$A_1$ to be reduced and that of $G$ have opposite signs in all cases except for
$D(2|n)$, cf. Appendix A. Thus in most cases
$c_{ESA}\geq 0\Leftrightarrow K_{cl}\geq 0$. This property turns out to hold in
the quantum case as well. In the case of $D(2|n)$, $\WG$ splits into two
commuting algebras having opposite sign central extensions\cite{BoFrLi}.
\vglue 0.6cm

{\bf\noindent 3. The Quantum Hamiltonian Reduction}
\vglue 0.2cm
{\it\noindent 3.1 BRST formalism}
\vglue 0.1cm
\baselineskip=14pt
In this section we discuss the quantum hamiltonian reduction \cite{BeOo}
for an affine Lie superalgebra $\HBG$ at level $k$, generated by
$J_{\A}(z)$ ($\A\in\DE^{0}$), $j_{\G}(z)$ ($\G\in\DE^{1}$)
and $H^{i}(z)$ ($i=1, \ldots, n$).
Let $T_{WZNW}(z)$ be the energy-momentum tensor of an affine Lie
superalgebra $\HBG$, defined by the Sugawara form:
\eqn
T_{WZNW}
&=&{1\over 2(k+h^{\vee})}
\left( :\sum_{\A\in\DE^{0}_{+}}{\A^{2}\over2}(J_{\A}J_{-\A}+J_{-\A}J_{\A})
              +\sum_{i=1}^{n}H^{i}H^{i}  \right.\CR
&+&\left.\sum_{\G\in\DE^{1}_{+}}(j_{\G}j_{-\G}-j_{-\G}j_{\G}):\right)
\label{eq:sug}
\enn
where $h^{\vee}$ is the dual Coxeter number of $\BG$ and  $: \ :$ denotes
the normal ordering.
In order to impose the constraint for the currents at the quantum
level, we have to ``improve" the energy-momentum tensor by a contribution
from the Cartan currents $H^{i}(z)$:
\eq
T_{improved}(z)=T_{WZNW}(z)-\mu\cdot\pa H(z).
\label{eq:imp}
\en
Here $\mu$ is an $n$-vector.
With respect to the improved energy-momentum tensor $T_{improved}(z)$,
the currents corresponding to the roots $\A$ have conformal
weights $1+\mu\cdot\A$.
We are concerned with a class of Lie superalgebras whose even subalgebras
are $G \oplus A_{1}$, where $G$ is a semisimple Lie algebra.
In the previous section we have considered the constraint $J_{-\T}(z) = 1$
but the currents have conformal dimension one, so in order for the
constraint to make sense,  $J_{-\T}(z)$
must have improved conformal dimension, 0, whereas the currents of $\WG$ should
continue to have conformal dimension 1.
This means that  the vector $\mu$ should satisfy
\eq
\mu\cdot \T = 1, \quad
\mu\cdot \A = 0, \quad \mbox{for $\A \in \DE (G) $}.
\label{eq:mu}
\en
These equations \refeq{eq:mu} determine the vector $\mu$
uniquely:
\eq
\mu={ \T \over \T^{2}}.
\en
{}From \refeq{eq:odd1}, we find that
the fermionic currents $j_{\G}(z)$ ($j_{-\G}(z)$) for the positive
roots $\G\in\DE^{1}_{+}$ have conformal weight 3/2 (1/2).
The current $J_{\T}(z)$ has conformal weight 2.

Now we use the BRST-gauge fixing procedure.
In the previous section we took the Drinfeld-Sokolov gauge
\refeq{eq:dsgauge} and derived the Poisson bracket structure
for the currents.
In order to study the representation of the algebra, it is
very useful to consider the free field representation.
This is realized in the classical case by taking the \lq\lq diagonal" gauge
\eqn
J_{\T}(z)&=&0 , \quad
\T\cdot H(z)= a_{0} \pa\phi(z),    \CR
j_{-\G}(z)&=& \sqrt{N_{-\G,-\T+\G}}\chi_{\G}(z),
\quad \mbox{for $\G\in\DE^{1}_{+}$}.
\enn
in \refeq{eq:current}.
In the quantum case we similarly introduce
a free boson $\phi(z)$, coupled to the
world sheet curvature, and
free fermions $\chi_{\G}(z)$, satisfying operator product expansions
\eq
\chi_{\G}(z)\chi_{\T-\G'}(w)={\D_{\G,\G'}\over z-w}+\cdots,
\quad \mbox{for $\G,\G' \in \DE^{1}_{+}$.}
\en
This is consistent with the OPE's of ${\bf \hat{g}}$ together with the
constraint. The role of the free fermions is to act as auxiliary fields that
convert the constraints into first class ones.

We further introduce fermionic ghosts ($b_{\T}(z)$,$c_{\T}(z)$) with
conformal weights (0,1) and bosonic ghosts
($\tilde{b}_{\G}(z)$, $\tilde{c}_{\G}(z)$) of weight
($\half$, $\half$) for $\G\in\DE^{1}_{+}$.
The BRST current $J_{BRST}(z)$  is defined as \cite{BeOo2}
\eq
J_{BRST}(z)=c_{\T} (J_{-\T}-1)+\sum_{\G\in\DE^{1}_{+}}
\tilde{c}_{\G} (j_{-\G}-\sqrt{N_{-\G,-\T+\G}}\chi_{\G})
+\half \sum_{\G\in\DE^{1}_{+}}
N_{\G,\T-\G}:\tilde{c}_{\G}\tilde{c}_{\T-\G}b_{\T}: .
\en
We can easily  show that the BRST charge $Q_{BRST}=\oint\dz J_{BRST}(z)$
satisfies the nilpotency condition $Q_{BRST}^{2}=0$.
The total energy-momentum tensor is expressed as
\eq
T_{total}(z)=T_{improved}(z)+T_{\chi}+T_{ghost}(z),
\en
where
\eqn
T_{ghost}(z)&=& :(\pa b_{\T})c_{\T}
                +\half \sum_{\G\in\DE^{1}_{+}}
               (\Tb_{\G} \pa \Tc_{\G} - (\pa \Tb_{\G}) \Tc_{\G}):, \CR
T_{\chi}(z)&=& -\half  \sum_{\G\in\DE^{1}_{+}}
      : \chi_{\T-\G}\pa \chi_{\G}:.
\enn
The central charge $c$ of the total system is computed to be
\eq
c_{total}=c_{WZNW}-12k\mu^{2}
 +\half |\DE^{1}_{+}|
 -2 -|\DE^{1}_{+}|,
\label{eq:cen}
\en
(where $|\DE^{1}_{+}|$ is still the number of positive odd roots).
The last two terms are contributions from ghost fields.
The central charge $c_{WZNW}$ of the WZNW models on the Lie
supergroup at level $k$ is given by the formula:
\eq
c_{WZNW}={k \ \sdim \BG \over k+h^{\vee}},
\en
where the super dimension $\sdim \BG$ of a Lie superalgebra $\BG$ is
defined as ${\rm dim} \BG_{0}-{\rm dim} \BG_{1}$.
The list for the corresponding central charges
is shown in table \ref{ta:ce}.
\begin{table}
\caption{Central charges for $G$ extended superconformal algebras}
\label{ta:ce}
\begin{center}
\begin{tabular}{|ll|}                                 \hline
$\BG$        &    $c_{total}$                          \\ \hline
$A(n|1)$     &    $[6k^{2}+k(n^{2}+3n-9)-n^{2}-2n+3]/[ k+n-1]$ \\
$B(n|1)$     &    $[(k+1)(4n^{2}+4n-15-6k)]/[ 2(k+3-2n)]$ \\
$D(n|1)$     &    $[(k+1)(4n^{2}-16-6k)]/[ 2(k+4-2n)]$  \\
$D(2|n)$     &    $[6 k^{2}+k(2n^{2}+3n-8)+4-4n^{2}]/[ k+2n-2]$  \\
$D(2|1;\A)$  &    $ -3-6k$ \\
$F(4)$       &    $[2(-2k^{2}+7k+9)]/[k-3]$    \\
$G(3)$       &    $[9k^{2}+13k-22]/[2(k+2)]$    \\ \hline
\end{tabular}
\end{center}
\end{table}
The results here are in complete agreement with
previous calculations using a variety of different techniques.
Thus, the results for $A(n|1), B(n|1), D(n|1)$ were obtained by Knizhnik and
Bershadsky \cite{KnBe}; the one for $D(2|1;\alpha )$ agrees with
original literature on the doubly extended $N=4$ algebra \cite{sigma,dob}
(see also our recent free field representation \cite{ItMaPe}),
the result for $D(2|n)$ and $G(3)$ agrees with
Bowcock and Fradkin and Linetsky \cite{BoFrLi},
and the result for $F(4)$ with that of Fradkin and Linetsky \cite{FrLi}.
Our treatment here, however, provides a unifying framework.

In the classical limit $k\rightarrow \infty$, the expression
\refeq{eq:ene} becomes $-12k/\T^{2}$, which agrees with the
result (\ref{eq:clope}) obtained in the previous section.
\vglue 0.4cm
{\it\noindent 3.2 The free field representation}
\vglue 0.1cm
\baselineskip=14pt
We consider the free field representations of $G$ extended superconformal
algebra based on the Wakimoto construction \cite{FeFr} of the affine Lie
superalgebra $\HBG$ \cite{Ito}.
Let us introduce bosonic ghosts $(\B_{\A}(z), \G_{\A}(z))$  for
even positive roots $\A\in\DE^{0}_{+}$ with conformal dimensions (1,0),
fermionic ghosts $(\eta_{\G}(z), \xi_{\G}(z))$ for odd positive
roots $\G\in\DE^{1}_{+}$ with (unimproved)
conformal dimensions (1,0) and $n$ free
bosons $\vp(z)=(\vp_{1}(z),\ldots,\vp_{n}(z))$ coupled to the world-sheet
curvature, satisfying the operator product expansions:
\eqn
\B_{\A}(z)\G_{\A'}(w)&=&{\D_{\A,\A'}\over z-w}+\cdots, \quad
\eta_{\G}(z)\xi_{\G'}(w)={\D_{\G,\G'}\over z-w}+\cdots, \CR
\vp_{i}(z)\vp_{j}(w)&=&-\D_{i j}{\rm ln}(z-w)+\cdots.
\enn
Using these free fields the energy-momentum tensor is expressed as
\eq
T_{WZNW}(z)=\sum_{\A\in\DE^{0}_{+}}:\B_{\A}\pa\G_{\A}:
           -\sum_{\G\in\DE^{1}_{+}}:\eta_{\G}\pa\xi_{\G}:
           -\half :(\pa\vp)^{2}:-{\I\rho\cdot\pa^{2}\vp \over \A_{+}},
\en
where $\A_{+}=\sqrt{k+h^{\vee}}$ and $\rho=\rho_{0}-\rho_{1}$.
$\rho_{0}$ ($\rho_{1}$) is half the sum of positive even (odd) roots:
\eq
\rho_{0}=\half\sum_{\A\in\DE^{0}_{+}}\A, \quad
\rho_{1}=\half\sum_{\G\in\DE^{1}_{+}}\G .
\en
 The Cartan part currents $\hat{H}^{i}(z)$ are given by
\eq
\hat{H}^{i}(z)=-\I\A_{+}\pa\vp^{i}
         +\sum_{\A\in\DE^{0}_{+}}\A^{i}:\G_{\A}\B_{\A}:
         +\sum_{\G\in\DE^{1}_{+}}\G^{i}:\xi_{\G}\eta_{\G}:.
\en
The currents indicated by a ``hat" will differ by the final full ones by
certain fermionic contributions to be described.
After improvement of the energy momentum tensor, the ghost systems acquire
conformal dimensions, $(0,1)$ for $(\beta_\theta,\gamma_\theta )$ and
$(\half ,\half)$ for $(\eta_\gamma ,\xi_\gamma )$.
In the BRST gauge fixing procedure it is natural to assume that the
multiplets $(\B_{\T}, \G_{\T}, b_{\T}, c_{\T})$ and
$(\eta_{\G}, \xi_{\G}, \tilde{b}_{\G}, \tilde{c}_{\G})$
for $\G\in\DE^{1}_{+}$ form  Kugo-Ojima quartets \cite{KuOj}, and that the
corresponding energy-momentum tensor is BRST-exact.
In the case of the $\widehat{sl(N)}$ affine Lie algebra, this has been
proven both  by  homological techniques \cite{Figo}, and by a more direct
method \cite{Haya}.
This ansatz enables us to write the total energy-momentum tensor
in the form:
\eq
T_{total}(z)=T_{ESA}(z)+\{ Q_{BRST}, *\},
\en
where $T_{ESA}$ is the energy-momentum tensor of $G$ extended
superconformal algebra:
\eqn
T_{ESA}(z)
&=& -\half :(\pa\vp)^{2}:
    -\I \left({\rho \over \A_{+}}-\A_{+}\mu\right)\cdot\pa^{2}\vp \CR
    &&+\sum_{\A\in\DE_{+}(G)}:\B_{\A}\pa\G_{\A}:
 +\half \sum_{\G\in\DE^{1}_{+}}:(\pa\chi_{\G})\chi_{\T-\G}: .
\label{eq:ene}
\enn
The central charge of $T_{ESA}$ is given by the formula
\eq
c=n-12 \left({\rho \over \A_{+}}-\A_{+}\mu\right)^{2}
  +2|\DE_{+}(G)|+\half |\DE^{1}_{+}|.
\label{eq:ccemg}
\en
This may be shown to be equal to $c_{total}$ \refeq{eq:cen}.

In the case that $\WG$ is expressed as a direct sum of simple
affine Lie algebras $\oplus_{i} \WG_{i}$,
the relation between the level $k$ of the affine Lie superalgebra $\HBG$
and that, $K_{i}$, of the $i$'th affine Lie algebra $\widehat{G_{i}}$, is
given by considering the decomposition of the vector
$\rho/\A_{+}-\A_{+}\mu$ into the roots of the even subalgebras
$(\oplus_{i}G_{i} )\oplus A_{1}$:
\eq
{\rho \over \A_{+}}-\A_{+}\mu
=\sum_{i}{\rho_{G_{i}}\over \A_{+}}
 +({\half-{|\DE^{1}_{+}|\over 4}\over \A_{+}}-{\A_{+}\over\T^{2}})\T.
\en
Here $\rho_{G_{i}}$ is half the sum of positive roots of the
subalgebra $G_{i}$.
The above decomposition means that
\eq
k+h^{\vee}={\A_{L i}^{2}\over 2}(K_{i}+H^{\vee}_{i}),
\label{eq:levels}
\en
where $H^{\vee}_{i}$ is the dual Coxeter number of the even
subalgebra $G_{i}$ in a notation appropriate to the long root being normalized
to length squared, 2, whereas $\A_{L i}$ is an actual long long root of the
subalgebra
$G_{i}$. Notice that the central extensions of the affine sub algebras
receive additional contributions from the free fermions, see next section.

The $n$ bosons $\vp(z)=(\vp_{1}(z), \ldots, \vp_{n}(z))$ coupled to the
world sheet curvature, can be divided into two classes, due to the
decomposition of the Cartan subalgebra $\Bh=H_{G}\oplus H_{A_{1}}$,
where $H_{G}$ and $H_{A_{1}}$ are  the Cartan subalgebras of the even
subalgebras $G$ and $A_{1}$, respectively.
A boson $\theta\cdot\vp$ in the $\theta$ direction of the root
space of the even subalgebra $A_{1}$, commutes with the bosons lying
along the root space of the even subalgebra $G$.
The remaining $n-1$ free bosons are used for the free field representation
of the $\hat{G}$ affine Lie algebra, combined with
($\B_{\A}$, $\G_{\A}$)-systems ($\A\in\DE_{+}(G)$).
If we define a Feigin-Fuchs boson $\phi$:
\eq
\phi (z)={\T\cdot\vp(z)\over \sqrt{\T^{2}}},
\en
the energy-momentum tensor \refeq{eq:ene} becomes
$T_{ESA}(z)=T_{\phi}(z)+T_{G}(z)+T_{\chi}(z)$, where
\eq
T_{\phi}=-\half (\pa\phi)^{2}-{\I Q\over 2} \pa^{2}\phi,
\quad
Q=\sqrt{\T^{2}}
({2-|\DE^{1}_{+}|\over 2\A_{+}}-{2\A_{+}\over\T^{2}}).
\en
$T_{G}(z)$ is the Sugawara energy-momentum tensor of the
affine Lie algebra $\WG$.

\vglue 0.6cm

{\bf\noindent 4. Generic Free Field Representations and Explicit Form of
all Algebras}
\vglue 0.2cm
Having provided a free field representation for the energy momentum tensor,
we still have to do the same for the affine $G$-currents and in particular
for the $|\DE^1_+|$ supercurrents, $G_{\G}$. We have at our disposal already
the free (Feigin Fuchs) scalar, $\phi(z)$ and the $|\DE^1_+|$
free fermions, $\chi_{\G}(z)$.
We then build the affine currents $J_{\A}(z),H_i(z)$ as
\bea
J_{\A}(z)&=&\hat{J}_{\A}(z)+J^f_{\A}(z)\nn
H_i(z)&=&\hat{H}_i(z)+H^f_i(z)
\eea
where the fermionic parts are given by
\bea
J^f_{\A}(z)&=&\hf\sum_{\G,\G'\in\DE^1_+}t^{\A}_{\G',\G}
:\chi_{\G}(z)\chi_{\T-\G'}(z):\nn
H^f_i(z)&=&\hf\sum_{\G\in\DE^1_+}(\G_{\perp})_i:\chi_{\G}(z)\chi_{\T-\G}(z):
\eea
The ``hatted" currents commute with the fermionic ones but are also affine
$G$ currents with levels such that the total match up to what is required.
More precisely, for $G=\oplus_i G^{(i)}$, the central extensions of the affine
subalgebras are denoted by $K^{tot}_i$ in a notation pertaining to the length
squared of
the maximal root being 2. We then have
\bea
K^{tot}_i&=&K_i+K^f_i\nn
K^f_i&=&\frac{c_F^{(i)}}{(\A^2_L)_i}
\eea
where $K_i$ is the contribution from the hat algebra tabulated in
table (\ref{ta:cis}).
These may be treated by the standard Wakimoto like free field
realisations\cite{FeFr} as above.

So far we have used the parameter, $k$, the central extension of the Lie
superalgebra  given by the ``hat'ed" currents, for expressing everything.
It may be convenient for various purposes and comparisons to express the
central  charges as functions of the central extensions of the $\WG$
algebra in a standard normalization pertaining to the longest root having
length squared equal to 2. This we do in table \ref{ta:cvsk}.
{}For $D(2|n)$ we have written the central charge in terms of the level of
the affine $sp(2n)$, and for $D(2|1;\alpha)$ we have written it
in terms of the level of $su(2)^+$, one of two commuting $su(2)$'s. The
parameter, $\G$ (not to be confused with the odd roots!) in the tables is
related to the continuous parameter, $\A$, occurring in $D(2|1;\A)$ as
$$\G\equiv\frac{\A}{1+\A}.$$

\begin{table}
\caption{Central charges for $G$ extended superconformal algebras, in terms of
the level $K^{tot}$ of the full affine algebra.}
\label{ta:cvsk}
\begin{center}
\begin{tabular}{|lll|}                                 \hline
$\BG$        &   k   &   $c_{total}$                          \\ \hline
$A(n|1)$     & $K^{tot}+1$
& $\frac{6(K^{tot})^{2}+K^{tot}(n^{2}+3n+3)+n}{ K^{tot}+n}$ \\
$B(n|1)$     & $-K^{tot}-1$
& $\frac{K^{tot}(6K^{tot}+4n^{2}+4n-9)}{ 2(K^{tot}+2n-2)}$ \\
$D(n|1)$     & $-K^{tot}-1$
& $\frac{K^{tot}(6K^{tot}+4n^{2}-10)}{ 2(K^{tot}+2n-3)}$  \\
$D(2|n)$     & $2K^{tot}_1$
& $\frac{12(K^{tot}_1)^2+K^{tot}_1(2n^2+3n-8)+2-2n^2}{K^{tot}_1+n-1}$\\
$D(2|1;\alpha)$  & $-\gamma(K^{tot}_1+1)$
& $-3+6\gamma(K^{tot}_1+1)$                  \\
$F(4)$       & $-K^{tot}-1$
& $\frac{2(2(K^{tot})^{2}+11K^{tot})}{K^{tot}+4}$    \\
$G(3)$       & $K^{tot}+1$
& $\frac{9(K^{tot})^{2}+31K^{tot}}{2(K^{tot}+3)}$  \\ \hline
\end{tabular}
\end{center}
\end{table}

Introducing the previously defined combinations \refeq{eq:JGGp}, we find for
the supercurrents in all  cases the following remarkably simple general
expression
\bea
G_{\G}(z)&=&\sum_{\G'\in\DE^1_+}
\{\D_{\G,\G'}[\frac{(k+\theta^2/2)}{\A_+}\pa+\frac{i}{\sqrt{2}}
\sqrt{\frac{\theta^2}{2}}\pa\phi(z)] \nn
&-&\frac{1}{\A_+}\sum_i
[\hat{J}^{(i)}_{\G,\G'}(z)+c^{(i)}_2J^{f,(i))}_{\G,\G'}(z)]\}\chi_{\G'}(z)
\eea
The only difference between the various different algebras is that the
constants, $c_1,c_2$ take different values. These are provided in table
(\ref{ta:cis}).
\begin{table}
\caption{Constants occurring in the free field representations.}
\label{ta:cis}
\begin{center}
\begin{tabular}{|l|rrrlrr|}                         \hline\hline
$\BG$        & $\theta^2/2$  & $c^{(1)}_2$ & $c^{(2)}_2$ & $k+h^\vee$
& $\frac{\A_L^2}{2}(K+H^{\vee})^{(1)}$
& $\frac{\A_L^2}{2}(K+H^{\vee})^{(2)}$                 \hl
$A(n|1)$     &  $-1$  & $1$ &$\frac{4(n+1)}{(n-1)^2}$&
$k+n-1 $& $K+n+1$ &   \\[3pt]
$B(n|1) $    &  $-2$  &               $0$ &&   $k-2n+3 $& $-(K+2n-1)$&     \\
$D(n|1)$     &  $-2$  &$               0$ &&   $k-2n+4 $& $-(K+2n-2)$&     \\
$D(2|n)$     &  $-1$  &$ 1$ &$ 0$ &$ k+2n-2$ &$ 2(K^{(1)}+n+1)$ &
$ -(K^{(2)}+2)$ \\
$D(2|1;\A)$  &$  -1$  &$ \frac{2}{3}$ &$ \frac{2}{3}$ &$ k$ &$-\G(K^{(1)}+2)$
&$- (1-\G)(K^{(2)}+2)$ \\
$F(4)$       & $\frac{3}{2}$ &     $\twt$ &&  $     k-3$ &$ -(K+5)  $ & \\
$G(3)$       &$-\frac{4}{3}$ &     $\twt$ &&  $     k+2$ &$   K+4   $ & \hl
\end{tabular}
\end{center}
\end{table}
\begin{table}
\caption{Coefficient constants in the general Extended Superconformal
Algebra.}
\label{ta:fis}
\begin{center}
\begin{tabular}{|l|rll|}                         \hline\hline
$\BG$        &       $  f_1/f_2^{(1)} $ & $f^{(1)}_2$ &$ f^{(2)}_2$ \hl
$A(n|1)$     & $k-1$& $\hf(k+h^{\vee}+k)$ &$ k-1$ \\[6pt]
$B(n|1)$     & $k+1$& $\hf(k+h^{\vee}+k+2)$ &  \\[6pt]
$D(n|1)$     & $k+1$& $\hf(k+h^{\vee}+k+2)$ &  \\[6pt]
$D(2|n)$     & $k+1$& $k+h^{\vee}+1$ & $k+1$  \\[6pt]
$D(2|1;\A) $ & $k+\G$& $k+(1-\G)$ & $k+\G$\\[6pt]
$F(4)$       & $k+1$ & $k+h^{\vee}+2$&  \\[6pt]
$G(3)  $     & $k-1$& $k+h^{\vee}-\frac{4}{3}$ & \hl
\end{tabular}
\end{center}
\end{table}
In the cases $A(n|1),D(2|n)$ and
$D(2,1;\A)$ the surviving affine algebra, $G$, is
nonsimple: $G=G^{(1)}\oplus G^{(2)}$, and the constants similarly break up
in two. These are listed separately. It is worth while noticing that these
constants are independent of the affine levels, expressed variously in
terms of the affine Lie super algebra level, $k$, or in terms of the
ordinary affine $G^{(i)}$ levels.

The above exceedingly compact form of the
generic free field realisation specializes to (much more complicated looking)
results in the special cases of $osp(N|2)$ and $A(n|1)$, $n>1$
\cite{Mat}, $A(1|1)$ \cite{Mats,ivanov}, and $D(2|1;\alpha)$ \cite{ivanov},
as well as our results for $D(2|n)$ and $D(2|1;\alpha)$ \cite{ItMaPe}.
The results for $F(4)$ and $G(3)$ are new.

The full quantum algebra may now be constructed directly from the free
field expressions. Indeed the constants are partly determined by demanding
closure. Most of the OPE's of the quantum ESA are rather trivial. Thus the
OPE's between the energy momentum tensor and the other generators merely
indicate that the conformal dimensions of the affine currents is 1, whereas
that of the supercurrents is $3/2$. Finally the energy momentum tensor with
itself requires knowledge of the central charges already provided in table
(\ref{ta:ce}).
Similarly the OPE's with the affine currents partly just indicate that the
energy momentum tensor is an affine scalar whereas the supercurrents are
primaries  transforming according to suitable representations as already
discussed at length. The affine currents with themselves generate the central
extensions of the affine algebras as wee have also discussed.

There remains the highly non trivial OPE's between the supercurrents
themselves. A glance at the several partly covered examples already presented
in the literature will convince anyone that previous formulations were
unsatisfactory. Now however, we are able to provide a very simple general
expression:
\bea
G_{\T-\G'}(z)G_{\G}(w)&=&\frac{f_1(k)}{k+h^{\vee}}
\frac{\D_{\G',\G}}{(z-w)^3}
-\frac{1}{k+h^{\vee}}\sum_i f_2^{(i)}(k)\left (
2\frac{J^{(i)}_{\G,\G'}(w)}{(z-w)^2}
+\frac{\pa J^{(i)}_{\G,\G'}(w)}{z-w}
\right )\nn
&-&\frac{2}{\T^2}\D_{\G,\G'}\frac{T(w)-T_G(w)}{z-w}
+\frac{1}{k+h^{\vee}}\frac{(J^2)_{\G,\G'}(w)}{z-w}
\label{eq:GG}
\eea
Here
$$T_G(w)\equiv\frac{1}{2(k+h^{\vee})}\{\sum_{\A\in\DE^0(G)}\frac{\A^2}{2}
J_{\A}J_{-\A}(w)+\sum_iH_iH_i(w)\}$$
is similar to the sugawara energy momentum tensor, but in fact carries the
incorrect normalization, namely the normalization that would be needed for
the ``hat" part whereas here we are only dealing with the full generators.
Also, the trace part of $(J^2)_{\G,\G'}(w)$
is of a similar form but (in general) with yet another normalization. In
each concrete case one may wish to combine these two contributions. The
great similarity between this expression and the classical one given
previously (and corresponding to the limit $k\rightarrow\infty$) is only
obtained when the composite operator $J^2$ is ``normal ordered" according to
the {\em symmetrized} prescription
$$:A(z)B(z):\equiv \hf\oint_z\frac{dw}{2\pi i}\frac{A(z)B(w)+A(w)B(z)}{z-w}.$$

It should be emphasized that the striking simplicity of these expressions is
perhaps slightly deceptive. This point is particularly relevant for the cases
where the algebra $G$ is non-simple. In these cases, namely, there is no \`a
priori natural relative normalization between the representation matrices and
indeed between the roots of the two commuting algebras, had it not been for
the fact that the structure arose from a Lie superalgebra reduction.
But the fact that these commuting algebras
are embedded in a {\em simple} Lie super algebra implies a precise
choice for that relative normalization. It is only with that very choice,
which allows us to introduce the currents $J_{\G,\G'}$ in a particular
normalization, that the simplicity obtains. So even though the final
expressions appear superficially to make no reference to the Lie super
algebra, in fact that underlying structure remains of importance.
\vglue 0.6cm

{\bf\noindent 5. Degenerate Representation of $G$ Extended Superconformal
Algebras}
\vglue 0.2cm
The free field realizations, which we have summarized for all extended
superconformal algebras above, can give rise to efficient calculational
methods. To begin with, however, they give rise to very highly reducible
representations. Thus, in order to discuss representation theory based on
them we must introduce appropriate screening charges and understand how
reducible representations can be viewed as cohomologies of those.
We here take a number of steps in this direction.

The Fock space of the $G$ extended superconformal algebras is a tensor
product of ones for
$|\DE^{1}_{+}|$ fermions $\chi_{\G}$, free fields for the affine Lie
algebra $\WG$, and a free boson $\phi(z)$ coupled to the world
sheet curvature.

The free field representations of affine Lie algebras $\WG$
follows from our discussion above\cite{FeFr}.
Let $\bA_{i}$ ($i=1,\ldots, r$) be simple roots of the even subalgebra $G$.
$\blm_{i}$ ($i=1, \ldots, r$) the fundamental weights of $G$ satisfying
${2\blm_{i}\cdot\bA_{j} \over \bA_{j}^{2}}=\D_{i j}$.
$\Phi^{\bLM}_{\blm}(z)$ is a primary field of the affine Lie algebra
$\widehat{G}$ at level $K$, with weight $\blm$ in the highest
weight module with highest weight $\bLM$.
In the free field representation, this field can be expressed
as $p^{\bLM}_{\blm}(z)\EXP ({\I\bLM\cdot\vp(z)\over\A_{+}})$, where
$p^{\bLM}_{\blm}$ is a polynomial consisting of terms,
$\G_{\A_{1}}\cdots \G_{\A_{k}}(z)$ ($\A_{i}\in\DE_{+}(G)$)
such that $\blm=-\bLM+\A_{1}+\cdots+\A_{k}$.
(Note that in the present prescription, the vertex operator
$\EXP ({\I\bLM\cdot\vp(z)\over\A_{+}})$ represents the lowest weight state
$\Phi^{\bLM}_{-\bLM}(z)$.)

Denote the total Fock space as
$F_{\chi,\bLM,p}=F^{\chi}\otimes F^{G}_{\bLM}\otimes F^{\phi}_{p}$,
where $F^{\chi}$ is a fermionic Fock space constructed from $\chi_{\G}$
($\G\in\DE^{1}_{+}$).
$F^{G}_{\bLM}$ is a Fock space of the algebra $\widehat{G}$ built on a
primary field $\Phi^{\bLM}_{-\bLM}=\E^{{\I\bLM\cdot\vp(z)\over \A_{+}}}$.
$F^{\phi}_{p}$ is a Fock space built on a vertex operator
$V_{p}(z)=\E^{\I p\sqrt{\T^{2}}\phi(z)}$.
The dual spaces $(F^{G}_{\bLM})^{*}$ and $(F^{\phi}_{p})^{*}$ are isomorphic
to $F^{G}_{-2\rho_{G}-\bLM}$ and $F^{\phi}_{-Q-p}$, respectively.

A primary field of a $G$ extended superconformal algebra is expressed
as the products of three fields:
\eq
V_{\G_{1}, \ldots, \G_{l}}{}^{\LM}_{\lm}{}_{p}(z)
=\chi_{\G_{1}}(z)\cdots\chi_{\G_{l}}(z)\Phi^{\LM}_{\lm}(z)
 \E^{\I p\sqrt{\T^{2}}\phi(z)},
\label{eq:prim}
\en
where $\G_{i}$ are positive odd roots.
The conformal weight of \refeq{eq:prim} is given by
\eq
\DE={l\over2}+{\LM(\LM+2\rho_{G})\over 2\A_{+}^{2}}
+\half (p^{2}+Q p)\T^{2}.
\label{eq:hesa}
\en
\vglue 0.4cm
{\it\noindent 5.1 Screening operators}
\vglue 0.1cm
\baselineskip=14pt
In order to study the representation of the algebra using free fields,
we must specify the screening operators which commute with the generators
of the extended superconformal algebra.
We consider screening operators which correspond to the simple
roots of the Lie superalgebra $\BG$.
These screening operators are BRST-equivalent to those of the
affine Lie superalgebra $\HBG$ \cite{BeOo}.
In the present choice of the simple root system of the Lie superalgebra in
table \ref{ta:li}, the simple roots of the even subalgebra $G$ are a
subset of those of $\BG$ (see Appendix A).
Thus, first of all we shall get the screening operators corresponding to
the simple roots of the affine Lie algebra $\WG$.
Since the remaining simple roots are odd,
they will correspond to fermionic type screening operators. As discussed
for example by Kato and Matsuda\cite{KaMa} fermionic screening operators
may only occur in single contour integrals without producing ambiguous
short distance behaviours. Hence it is convenient to introduce yet another
screening operator, which characterizes
the $A_{1}$ even subalgebra corresponding to the root $\T$, even though that
root does not belong to our choice of simple roots.
\vglue 0.4cm
{\it\noindent 5.1.1. Affine screening operators}
\vglue 0.1cm
\baselineskip=14pt
First we can take the standard
screening operators $S_{\bA_{i}}(z)$ of the affine
Lie algebra $\widehat{G}$ as those of $G$ extended superconformal algebra:
\eq
S_{\bA_{i}}(z)=s_{\bA_{i}}(z)\E^{-{\I\bA_{i}\cdot\vp(z)\over\A_{+}}},
\label{eq:scr1}
\en
where $s_{\bA_{i}}$ consists of terms like $\B_{\bA_{i}}$  and
$\G_{\A_{1}}\cdots \G_{\A_{k}}\B_{\A_{1}+\cdots+\A_{k}+\bA_{i}}$
with $\A_{1}, \ldots, \A_{k}\in\DE_{+}(G)$.
These screening operators are used for the characterization of
singular vectors in the Fock modules of the affine Lie algebra
$\WG$.
\vglue 0.4cm
{\it\noindent 5.1.2. Fermionic screening operators}
\vglue 0.1cm
\baselineskip=14pt
Next, we consider the screening operators, which
corresponds to the odd simple roots. In the case of $A(n|1)$ there are two
of those, in all other cases there is only one, cf. Appendix A.
We have a fermion for every (negative) odd root (our labeling pertains to the
positive odd roots),
corresponding to the fundamental representation of the even subalgebra
$G$ of dimensions $|\DE^{1}_{+}|$.
Denote the highest weight of these representations as
$\bLM^{*}$. As we have seen, the odd roots themselves (suitably projected
to the root subspaces of the appropriate even sub algebras) may be taken as
the weights.
In order that the screening operator commutes with the
$\WG$ currents, it should be a $\WG$ singlet operator {\it i.e.}
the operator product expansion with the $\WG$ currents should be
regular.
These observations lead to fermionic screening operators of
the following type:
\eq
S_{f}(z)= \sum_{\G}\chi_{\G}\Phi^{\bLM^{*}}_{\T-\G}
          \EXP(-{\I\sqrt{\T^{2}}\phi(z)\over 2\A_{+}}),
\label{eq:scr2}
\en
where $\G$ runs over the roots corresponding to the weights of the
representation (with the highest weight $\bLM^{*}$).

The case of $A(n|1)$, where the algebra $G$ is $sl(n+1)\oplus u(1)$,
is slightly more complicated than the other cases due to the fact that
the representation of the fermions (or the supercurrents) is reducible:
Roots $\G$ and $\T-\G$ belong to different irreducible representations.
Correspondingly we get two odd roots and two fermionic screening operators by
restricting the sum above to weights of one or the other irreducible
representation. Further, it may be convenient to explicitly bosonize the
(``hat part" of the) $u(1)$ algebra. We refer to our paper\cite{ItMaPe}
for more details.

One may verify that these fermionic screening operators indeed either
commute with or produce total derivatives in OPE's with the generators.
{}For the supercurrents this holds by virtue of the Knizhnik-Zamolodchikov
equations \cite{KnZa}.

This treatment generalizes previous ones in special
cases\cite{Mat,Mats,Miki,ItMaPe} and again provides a unified account.
\vglue 0.4cm
{\it\noindent 5.1.3. Screening operators corresponding to
the $\theta$-direction}
\vglue 0.1cm
\baselineskip=14pt
Finally we need a screening operator to characterize
the $\T$ direction.
Denote this screening operator by $S_{\T}(z)$.
Based on several worked out examples we expect that this operator takes
the form:
\eq
S_{\T}(z)=s_{\T}(z) \EXP( {2\I\A_{+}\phi(z)\over \sqrt{\T^{2}}}),
\label{eq:scr5}
\en
where $s_{\T}(z)$ is a $G$ singlet operator
containing (among many others) a term $\prod_{\G\in\DE^{1}_{+}}\chi_{\G}$, and
hence having conformal dimension $|\DE^{1}_{+}|/2$. The requirement that
$S_{\T}(z)$ has conformal dimension 1, then fixes the ``momentum" in the vertex
operator part.
This assumption is justified in part by the following list of results:

When we consider the $osp(N|2)$ case, for $N=1$ and $2$, the screening
operators are the well known ones of $N=1$ and $2$ minimal
models, and they are of this form: In fact they may be
given as $S_{\T}=\psi_{1}\EXP(-\A_{+}\phi(z))$ and
$S_{\T}=(\psi_{1}\psi_{2}-{1\over K}J_{12})\EXP(-\A_{+}\phi(z))$,
respectively.

{}For the $N=3$ case, this kind of screening operator has been found by
Miki \cite{Miki}.

In the case of $N=4$ $sl(2)$ superconformal algebra this
screening operator has been obtained by Matsuda \cite{Mats}, and it is again
of the above form. Also we\cite{ItMaPe} have recently provided the
explicit form for the so-called doubly extended $N=4$ algebra\cite{sigma,dob}
corresponding to the case of $D(2|1;\A)$ which generalizes the two
$N=4$ cases corresponding to affine $so(4)$ and $su(2)$.

One must expect that this procedure can be generalized to any $N$, but so far
we have not found completely general expressions. From the cases studied it
seems clear that we must expect all kinds of terms with pairs of fermions from
$\prod_{\G\in\DE^{1}_{+}}\chi_{\G}$ being replaced by appropriate affine
current operators, in such a way that the combined group theory quantum numbers
work out correctly.

In the following we assume the existence of this kind of screening
operator for any $G$. Even though we do not have the explicit form of
$s_{\T}(z)$ in all cases, the knowledge of the quantum numbers suffices for the
determination of singular vectors as we shall see.
\vglue 0.4cm
{\it\noindent 5.2. Structure of null fields}
\vglue 0.1cm
\baselineskip=14pt
Based on the above observations on the screening operators, we discuss
the structure of singular vectors of $G$ extended superconformal algebras.
We consider the Neveu-Schwarz sector for simplicity.

Firstly we consider the singular fields corresponding to the affine Lie
algebra $\WG$ \cite{FeFr}.
These are given by the following screened vertex operators:
\eq
\Psi^{\bLM}_{\B_{1},\ldots, \B_{m}}(z)=\oint d u_{1} \cdots d u_{m}
          S_{\B_{1}}(u_{1})\cdots S_{\B_{m}}(u_{m})
          \Phi^{\bLM}_{-\bLM}(z),
\label{eq:singkm}
\en
where $\B_{a}$ ($a=1,\ldots, m$) are simple roots of the algebra $G$:
i.e. $\B_{a}=\bA_{i_{a}}$ for some $i_{a}=1, \ldots, r$.
The contours of integrations are taken for example
as in the prescription by Kato and Matsuda \cite{KaMa}.
The ``mass-shell-condition" that the above contour integral represents a
singular field is derived either by considerations of pole
singularities\cite{KaMa}, or
equivalently by noticing that the quantum numbers of the contour integral are
the same as those of the field $\Phi^{\bLM}_{-\bLM}(z)$ whereas, if the result
is a descendant it must be of a primary pertaining to the weight
$$\bLM-\sum_{a=1}^{m}\B_{a}.$$
It is then a simple matter to work out the difference in conformal dimension
of these two fields, using \refeq{eq:hesa} and demand that it be an integer:
\eq
{1\over{\A_{+}^{2}}}\sum_{a<b} \B_{a}\cdot\B_{b}
-{1\over\A_{+}^{2}}\sum_{a=1}^{m}\B_{a}\cdot\bLM = -M,
\label{eq:onkm}
\en
with a positive integer $M$.
If $\sum_{a=1}^{m}\B_{a}=n'\bA$ and $M=n n'$ for a positive root
$\bA\in\DE_{+}(G)$ and positive integers $n$ and $n'$, we get the
Kac-Kazhdan formula \cite{KaKa}:
\eq
(\bLM+\rho_{G})\cdot \bA=+n\A_{+}^{2}+ n' {\bA^{2}\over2}.
\label{eq:kaka}
\en
This formula and its dual, which is obtained by replacing $\bLM$
by $-2\rho_{G}-\bLM$, characterize the singular vectors of the
Fock module of the affine Lie algebra $\WG$.

The fermionic singular vectors are given by considering a screened
vertex operator of the form:
\eq
\Psi^{\bLM,p}(z)=\oint d u S_{f}(u)V_{\bLM,p}(z),
\en
where $V_{\bLM,p}(z)=\Phi^{\bLM}_{\blm}\EXP(\I p\sqrt{\T^{2}}\phi(z))$.
again the contour integral has the quantum numbers of $V_{\bLM,p}(z)$, but
(if singular) must be a descendant of a primary of the Fock module
$F^{G}_{\bLM+\bLM^{*}}\otimes F^{\phi}_{p-{1\over 2\A_{+}}}$
since $S_f$ carries momentum $-1/2\A_+$.
Equivalently the non-zero existence of the above contour integral requires
the condition:
\eq
{\bLM\cdot \bLM^{*} \over \A_{+}^{2}}-{p\T^{2}\over 2\A_{+}}=-M,
\label{eq:fersi}
\en
where $M$ is a positive integer. In this case $\Psi^{\LM,p}(z)$ is
a singular vector at level $M-\half$.

The null fields in the $\T$ direction can be obtained from the
screened vertex operators:
\eq
\Psi^{p}_{r}(z)=\oint d u_{1}\cdots d u_{r}
              S_{\T}(u_{1})\cdots S_{\T}(u_{r}) V_{p}(z),
\en
where $V_{p}(z)=\E^{\I p\sqrt{\T^{2}}\phi (z)}$.
The on-shell condition becomes
\eq
{r(r-1)\over 2}({2\A_{+}\over \T^{2}})^{2}\T^{2}+2r p \A_{+}=-M.
\en
with an positive integer $M$.
Writing $M$ as ${r (s -|\DE^{1}_{+}|+2) \over 2}$, we find that
$p$ is given by
\eq
p_{r,-s}=-{r-1\over \T^{2}}\A_{+}-{s-|\DE^{1}_{+}|+2 \over 4\A_{+}}.
\en
In this case $\Psi^{p}_{r}(z)$ is a null field in the Fock module
$F^{\phi}_{-Q-p_{r,s}}$ at level ${r s\over 2}$, where
\eq
p_{r,s}={1-r\over \T^{2}}\A_{+}+{s+|\DE^{1}_{+}|-2 \over 4\A_{+}}.
\label{eq:virsi}
\en
The precise range for $s$ cannot be identified unambiguously until the
multiplicities of zeros of the Kac determinant (or something equivalent)
has been dealt with.

The formulas \refeq{eq:kaka}, \refeq{eq:fersi} and \refeq{eq:virsi}
characterize the whole singular vector structure of the $G$
extended superconformal algebras.
\vglue 0.6cm

{\bf\noindent 6.Conclusions and Discussion}
\vglue 0.2cm
In the present paper we have studied $G$ extended superconformal
algebras from the viewpoint of  classical and quantum hamiltonian
reductions of an affine Lie superalgebra $\HBG$, with even Lie
subalgebras $\WG\oplus \widehat{sl(2)}$. That framework seems to provide
an attractive classification scheme for extended superconformal algebras of the
type with generators consisting of just the energy-momentum tensor, the
(dimension $3/2$) supercurrents and some affine currents, the commutation
relations, however,
being allowed to contain quadratic composites of these generators.

We have derived very compact generic expressions for all these algebras and
their free field realizations in the classical case, and we have demonstrated
that these expressions also hold in the quantum case modulo simple
``renormalizations" of certain constants. We have provided all these constants,
thereby completing previous works with partial results
\cite{KnBe,BeOo,Bers,BeOo2,ItMaPe,BoFrLi,Mat,Mats,Miki}.

We have introduced a set of screening operators for all these $G$ extended
superconformal algebras.
Using the null field construction,
we have identified the primary fields corresponding to degenerate
representations of these algebras.

{}For future work, it would be of interest to work out in detail the
cohomologies
of the BRST operators we have constructed in order to understand the origin
of the free field realizations more completely from that point of view.
In general the quantum hamiltonian reduction which we have employed works
``algorithmically" only at the level of the energy momentum tensor. It
would be very useful to have a better understanding of how the remaining
generators are built in terms of free fields as well. The fact that we have
found expressions of such generic nature for the supercurrents, may perhaps
yield some clues towards such an understanding.

{}For the screening operators our understanding seems to be enough for
constructing singular vectors. Thus it would seem possible to find the
Kac-determinants  in all cases. For detailed calculation of correlation
functions there still remains the task of constructing the completely
explicit version of one of the screening  charges, which is so far known only
in a finite number of cases. With this knowledge
it is in principle possible to calculate correlation functions and
characters for these models. They will then be expressed as
products of those of $\WG$ affine Lie algebras, Virasoro
minimal models and free fermions.

Compared to the linearly extended superconformal algebras, a
geometrical interpretation of the present non-linear algebras
is unclear.
It seems an interesting problem to try to find a way of interpreting
these non-linear symmetries in terms of non-linear $\sigma$-models on
non-symmetric Riemannian manifolds.
This problem is important in order to clarify the geometrical
meaning of the $W$-algebras.

The present construction of the extended superconformal algebra
is based on the hamiltonian reduction of the affine Lie superalgebras.
It is well understood that in the case of $W$-algebras associated with
simple Lie algebras, the hamiltonian reduction
\cite{DrSo,Be,BeOo,BaFeFoRaWi} provides a connection to various
integrable systems such as Toda field theory \cite{BiGe} and
the generalized KdV hierarchy \cite{Ya}.
In the present case it is natural to expect the super Liouville model
coupled to Wess-Zumino-Novikov-Witten (WZNW) models or the KdV hierarchy
coupled to affine Lie algebras to arise.
In the bosonic case, this kind of integrable system has been partially
studied \cite{BaTjDr}.

One might further generalize the hamiltonian reduction procedure to
any Lie superalgebra. This would then give rise to a super $W$-algebra
coupled to WZNW models. Interesting work in this direction has been undertaken
in particular in the classical case\cite{FrRaSo}.
\vglue 0.6cm
{\bf Acknowlegement:} This work was supported in part by EEC contract
no. SC1 0394 C (EDB).
\vglue 0.6cm

{\bf\noindent Appendix. Root Systems of Lie Superalgebras}
\vglue  0.2cm
In this appendix, we describe the root systems of the Lie
superalgebras  with the even subalgebra $G\oplus A_{1}$ as given in
table \ref{ta:li} (with the exception of $B(1|n)$, cf. the discussion in the
text).
$\T$ is the simple root of $A_{1}$.
We use the orthonormal basis $e_{i}$ ($i\geq 1$) with positive
metric and $\D_{j}$ ($j\geq 1$) with negative metric:
\eq
e_{i}\cdot e_{j}=\D_{i j}, \quad
\D_{i}\cdot\D_{j}=-\D_{i j}, \quad
e_{i}\cdot \D_{j}=0.
\en
\begin{enumerate}
\item $A(n|1)$ $(n\geq 1)$,
(rank $n+2$,  the dual Coxeter number $h^{\vee}=n-1$) \\
Simple roots:

$\A_{1}=\D_{1}-e_{1}$,   \quad
$\A_{i}=e_{i-1}-e_{i}$, ($i=2, \ldots, n+1$), \quad
$\A_{n+2}=e_{n+1}-\D_{2}$.  \\
Positive even roots:
$e_{i}-e_{j}$, ($1\leq i<j\leq n+1$),  \quad
$\theta=\D_{1}-\D_{2}$. \\
Positive odd roots:
$\D_{1}-e_{j}$,  \quad $e_{j}-\D_{2}$, ($j=1,\ldots, n+1$).
\item $B(n|1)$ $(n\geq 0)$,
 (rank $n+1$,  $h^{\vee}=3-2n$)\\
Simple roots:

$\A_{1}=e_{1}-\D_{1}$, \quad
$\A_{i+1}=\D_{i}-\D_{i+1}$,  ($i=1, \ldots, n-1$), \quad
$\A_{n+1}=\D_{n}$. \\
Positive even roots:
$\theta=2e_{1}$, \quad
$\D_{i}\pm\D_{j}$,   ($1\leq i < j \leq n$),
$\D_{i}$,  ($i=1, \ldots, n$)  \\
Positive odd roots:
$e_{1}$, \quad
$e_{1}\pm\D_{j}$,
($j=1,\ldots, n$).
\item $D(n|1)$ $(n\geq 2)$,
 (rank $n+1$, $h^{\vee}=4-2n$)\\
Simple roots:

$\A_{1}=e_{1}-\D_{1}$,   \quad
$\A_{i+1}=\D_{i}-\D_{i+1}$,  ($i=1, \ldots, n-1$), \quad
$\A_{n+1}=\D_{n-1}+\D_{n}$. \\
Positive even roots:
$\theta=2e_{1}$, \quad
$\D_{i}\pm\D_{j}$,   ($1\leq i < j \leq n$). \\
Positive odd roots:
$e_{1}\pm\D_{j}$, \quad  ($j=1,\ldots, n$).
\item  $D(2|n)$ $(n\geq 1)$
 (rank $n+2$, $h^{\vee}=2n-2$\\
Simple roots:

$\A_{1}=-\D_{2}-\D_{1}$, \quad
$\A_{2}=\D_{1}-e_{1}$, \quad
$\A_{i+2}=e_{i}-e_{i+1}$,  ($i=1, \ldots, n-1$), \quad
$\A_{n+2}=2e_{n}$. \\
Positive even roots:
$-\D_{2}-\D_{1}$,
$\theta=\D_{1}-\D_{2}$, \quad
$e_{i}+ e_{j}$, ($1\leq i \leq j \leq n$),
$e_{i}- e_{j}$, ($1\leq i < j \leq n$). \\
Positive odd roots:
$\D_{1}\pm e_{j}$,
$-\D_{2}\pm e_{j}$,
($j=1,\ldots, n$).
\item $D(2| 1;\A)$ ($\A\neq 0, -1, \infty$),
 (rank  3, $h^{\vee}=0$)\\
Simple roots:

$\A_{1}=\half(\sqrt{2\G}\D_{1}+\sqrt{2(1-\G)}\D_{2}+\sqrt{2}e_{3})$,
\quad
$\A_{2}=-\sqrt{2\G}\D_{1}$,  \quad
$\A_{3}=-\sqrt{2(1-\G)}\D_{2}$,
where $\A=\G/(1-\G)$. \\
Positive even roots:
$\A_{2}$, \quad
$\A_{3}$, \quad
$\theta=2\A_{1}+\A_{2}+\A_{3}$. \\
Positive odd roots:
$\A_{1}$, \quad
$\A_{1}+\A_{2}$, \quad
$\A_{1}+\A_{3}$, \quad
$\A_{1}+\A_{2}+\A_{3}$.
\item  $F(4)$
(rank:$4$, $h^{\vee}= -3$)\\
Simple roots:

$\A_{1} = {1\over 2}(\sqrt{3}e_{1}+\D_{1}+\D_{2}+\D_{3})$, \quad
$\A_{2} = -\D_{1}$,  \quad
$\A_{3} = \D_{1}-\D_{2}$,  \quad
$\A_{4} = \D_{2}-\D_{3}$. \\
Positive even roots:
$\A_{2}$,\quad
$\A_{3}$,  \quad
$\A_{4}$,  \quad
$\A_{2}+\A_{3}$, \quad
$\A_{3}+\A_{4}$, \quad
$2\A_{2}+\A_{3}$, \quad
$\A_{2}+\A_{3}+\A_{4}$,
$2\A_{2}+\A_{3}+\A_{4}$, \quad
$2\A_{2}+2\A_{3}+\A_{4}$, \quad
$\theta=2\A_{1}+3\A_{2}+2\A_{3}+\A_{4}$. \\
Positive odd roots:
$\A_{1}$, \quad
$\A_{1}+\A_{2}$, \quad
$\A_{1}+\A_{2}+\A_{3}$, \quad
$\A_{1}+2\A_{2}+\A_{3}$, \quad
$\A_{1}+\A_{2}+\A_{3}+\A_{4}$, \quad
$\A_{1}+2\A_{2}+\A_{3}+\A_{4}$,  \quad
$\A_{1}+2\A_{2}+2\A_{3}+\A_{4}$, \quad
$\A_{1}+3\A_{2}+2\A_{3}+\A_{4}$.
\item $G(3)$
(rank:$3$,  $h^{\vee}=2$)\\
Simple roots:

$\A_{1}=\sqrt{{2\over3}}\D_{1}+{2e_{1}-e_{2}-e_{3}\over 3}$, \quad
$\A_{2}={-e_{1}+2e_{2}-e_{3}\over 3}$,\quad
$\A_{3}=-e_{2}+e_{3}$. \\
Positive even roots:
$\A_{2}$, \
$\A_{3}$, \
$\A_{2}+\A_{3}$, \
$2\A_{2}+\A_{3}$, \
$3\A_{2}+\A_{3}$, \
$3\A_{2}+2\A_{3}$, \
$\theta=2\A_{1}+4\A_{2}+2\A_{3}$. \\
Positive odd roots:
$\A_{1}$, \quad
$\A_{1}+\A_{2}$, \quad
$\A_{1}+\A_{2}+\A_{3}$,\quad
$\A_{1}+2\A_{2}+\A_{3}$, \quad
$\A_{1}+3\A_{2}+\A_{3}$, \quad
$\A_{1}+3\A_{2}+2\A_{3}$, \quad
$\A_{1}+4\A_{2}+2\A_{3}$.
\end{enumerate}
\vglue 0.8cm

{\bf References}
\vglue 0.4cm

\end{document}